\documentclass[twocolumn,english,english,aip,jcp,reprint]{revtex4-1}
\usepackage[T1]{fontenc}
\usepackage[latin9]{inputenc}
\usepackage{graphicx}

\makeatletter
 
 \@ifundefined{textcolor}{}
 {%
   \definecolor{BLACK}{gray}{0}
   \definecolor{WHITE}{gray}{1}
   \definecolor{RED}{rgb}{1,0,0}
   \definecolor{GREEN}{rgb}{0,1,0}
   \definecolor{BLUE}{rgb}{0,0,1}
   \definecolor{CYAN}{cmyk}{1,0,0,0}
   \definecolor{MAGENTA}{cmyk}{0,1,0,0}
   \definecolor{YELLOW}{cmyk}{0,0,1,0}
 }

\usepackage{subfigure}

\usepackage{babel}
\usepackage{braket}
\makeatother

\makeatother

\usepackage{babel}

\begin{document}

\title{Few simple rules governing hydrogenation of graphene dots}

\author{M. Bonfanti}

\affiliation{Dipartimento di Chimica Fisica ed Elettrochimica, Università degli
Studi di Milano, v. Golgi 19, 20133 Milan, Italy}

\author{S. Casolo}

\affiliation{Dipartimento di Chimica Fisica ed Elettrochimica, Università degli
Studi di Milano, v. Golgi 19, 20133 Milan, Italy}

\author{G. F. Tantardini}

\affiliation{Dipartimento di Chimica Fisica ed Elettrochimica, Università degli
Studi di Milano, v. Golgi 19, 20133 Milan, Italy}

\affiliation{Istituto di Scienze e Tecnologie Molecolari, Consiglio Nazionale
delle Ricerche, v. Golgi 19, 20133 Milano, Italy }

\affiliation{CIMaINa, Interdisciplinary Center of Nanostructured Materials and
Interfaces, v. Celoria 16, 20133 Milan, Italy. }

\author{A. Ponti}

\affiliation{Istituto di Scienze e Tecnologie Molecolari, Consiglio Nazionale
delle Ricerche, v. Golgi 19, 20133 Milano, Italy }

\author{R. Martinazzo}

\email{rocco.martinazzo@unimi.it}

\affiliation{Dipartimento di Chimica Fisica ed Elettrochimica, Università degli
Studi di Milano, v. Golgi 19, 20133 Milan, Italy}

\affiliation{Istituto di Scienze e Tecnologie Molecolari, Consiglio Nazionale
delle Ricerche, v. Golgi 19, 20133 Milano, Italy }

\affiliation{CIMaINa, Interdisciplinary Center of Nanostructured Materials and
Interfaces, v. Celoria 16, 20133 Milan, Italy. }
\begin{abstract}
We investigated binding of hydrogen atoms to small Polycyclic Aromatic
Hydrocarbons (PAHs) - \emph{i.e.} graphene dots with hydrogen-terminated
edges - using density functional theory and correlated wavefunction
techniques. We considered a number of PAHs with 3 to 7 hexagonal rings
and computed binding energies for most of the symmetry unique sites,
along with the minimum energy paths for significant cases. The chosen
PAHs are small enough to not present radical character at their edges,
yet show a clear preference for adsorption at the edge sites which
can be attributed to electronic effects. We show how the results,
as obtained at different level of theory, can be rationalized in detail
with the help of few simple concepts derivable from a tight-binding
model of the $\pi$ electrons. 
\end{abstract}
\maketitle

\section{Introduction}

Graphene, the recently discovered two-dimensional form of carbon\citep{novoselov04},
is a promising material for a future carbon-based nanoelectronics.
Its peculiar $\pi-\pi^{*}$ electronic band structure, with a linear
energy dispersion close to the Fermi level, introduces subtle quantum
pseudo-relativistic effects in the low-energy charge carrier dynamics
which hugely impact on the transport properties\citep{castroneto09,peres10,RossiRMP2011}.
This results, \emph{e.g.}, in a robust anomalous quantum Hall effect\citep{novoselov05,zhang05},
a universal conductivity minimum\citep{geim07} and ballistic transport
which can reach the micrometer scale\citep{schedin07}. From a practical
point of view, the substrate thickness, the high mobility of its charge
carriers and their (high-field) high saturation velocity represent
attractive features for the chip-makers. Nanostructuring, however,
is needed for applications, \emph{e.g.} for devising graphene-based
logic transistors where a band-gap is needed to achieve high operational
on-off ratios. Graphene Nanoribbons (GNRs) can be cut which show either
semiconducting or metallic properties, the latter coming with edge
states of unusual magnetic properties, possibly leading to carbon
based nanomagnets\citep{enoki07,Son2006}. Likewise, Graphene Dots
(GDs) can be designed to have specific electronic structures and transport
properties, by acting just on their shape and their connectivity.
GDs have been suggested for realizing spin qubits\citep{Burkard07},
spin filters\citep{Ezawa2008,GuoNanoflakes} and spin-logic devices\citep{KaxirsYayzev09},
and proposed as biomedical imaging agents\citep{bioimaging11} and
light absorbers for photovoltaics\citep{liang-shi10}. Transport properties
have been measured on a variety of dot devices carved entirely from
graphene by high-resolution electron-beam lithography\citep{GeimQD}. 

Most of these properties arise entirely from the $\pi$ electrons
and remain unaltered when saturation of the dangling $\sigma$ bonds
occurs, \emph{e.g.} in forming Polycyclic Aromatic Hydrocarbons (PAH).
The latter offer an enhanced chemical stability, and their nanostructuring
(energy level arrangement, interfacing with other materials, etc.)
can be realized with the help of well-developed carbon chemistry methods.
They have been used as building blocks for atomically-precise nanoribbon
fabrication\citep{Ruffieux10} and, in principle, may form the basis
for a bottom-up approach to realize arbitrarily complex carbon-nanostructures.
PAHs have also been investigated in many other fields, from petroleum
chemistry to astrochemistry. For instance, in the interstellar medium
(ISM), \emph{i.e.} the extremely rarefied medium which fills the space
between stars, the observed abundance of molecular hydrogen cannot
be explained by direct gas-phase routes involving $H$ atoms only,
rather is believed to occur on the carbonaceous surface of dust grains\citep{gould63,Hollenbach71}
and small carbonaceous particles. PAHs, which are estimated to lock
up \emph{ca.} 15\% of the interstellar carbon, have been suggested
as possible catalysts for H$_{2}$ formation \citep{Bauschlicher98,Fortes04,HornekaerAstro08}. 

In this work, we investigate the reaction of atomic hydrogen with
a number of PAHs, complementing previous related studies\citep{Bauschlicher98,zwier10,HornekaerAstro08,Ahlrichs00,hammer11}
which showed preference for addition at the edges of selected PAH
molecules. The main aim of this study was to emphasize the importance
of substrate relaxation ({}``geometrical'') effects in determining
a preference towards the edges. To this end, we selected substrate
PAH molecules with relatively small (sub-nanometer) dimensions, in
such a way to prevent any enhanced chemical reactivity at the edges
due to a true radical character (single occupation of a semilocalized
edge state), as it occurs for instance at the edges of wide zig-zag
GNRs. However, as we shall see in the following, some edge localization
is \emph{always} present. This provides an enhancement of the edge
reactivity which is of purely electronic origin and can be easily
understood in terms of few concepts derivable from a tight-binding
(H$\ddot{u}$ckel) model for the $\pi$ electrons. 

In addition, depending on the number of carbon atoms available for
the $\pi$ electron system and their connectivity, the systems considered
can also show a marked \emph{sub}lattice preference due to the {}``alternating
paths'' followed by (unpaired) itinerant electrons in graphenes (\emph{i.e.}
due to the presence of staggered midgap states). This is similar to
graphene\citep{casolo09,Intech11}, where these states form the basis
for a preferential sticking mechanism\citep{Hornekaer2006a,Rogeau2006}
forming \emph{para}-dimers (\emph{i.e.} two H atoms on opposite corners
of the same ring). We therefore distinguish two classes of PAHs according
to whether the number of sites in each sublattice is balanced or not,
and show how a final set of rules governing the site reactivity results
from the interplay of different electronic effects. The basic concepts
underlying these rules equally apply to larger systems, and thus allow
one to easily predict the chemical reactivity of $sp^{2}$ carbon
nanostructures with monovalent species forming covalent bonds with
the substrate. 

Importantly, in the present study we also take advantage of the modest
size of the systems investigated, and exploit the unique opportunity
of assessing the quality of the results of commonly used Density Functional
Theory (DFT) methods in investigating chemically-derived graphene
structures. This is done here by complementing the DFT data with those
obtained by using more accurate correlated wavefunction techniques.
\emph{En passant}, we briefly discuss the magnetic properties of pristine
and hydrogenated PAHs, which turn out to be well predicted by Lieb's
theorem\citep{TheoLieb}, in agreement with previous studies on triangularly
and hexagonally shaped GDs\citep{PalaciosPAH} and other defective
graphenic structures\citep{casolo09,Intech11}. 

The paper is organized as follows. Section I introduces some basic
properties of $\pi$-conjugated carbon systems which underlie the
presentation of the results given in Section III, after Section II
has provided the computational details of our calculations. Section
IV summarizes and concludes. 

Notice that in the following we adopt a surface science terminology,
whereby {}``adsorption to the substrate'' (here meant to be chemisorption)
is used interchangeably with {}``binding to the molecule''. %
\begin{figure}
\noindent \begin{centering}
\includegraphics[clip,width=0.9\columnwidth]{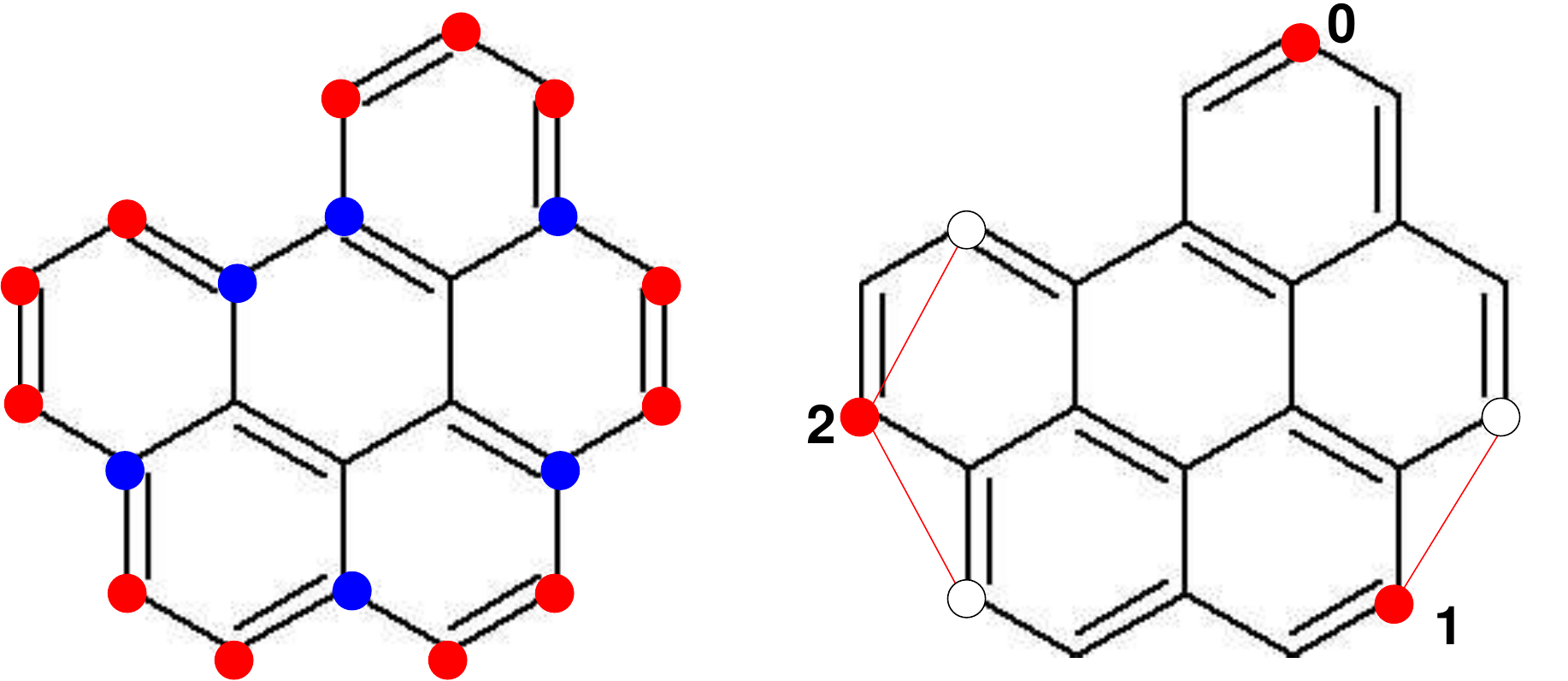}
\par\end{centering}

\caption{\label{fig:coordination}Left: Two- (red) and three- (blue) coordinated
edge sites in the benzo{[}ghi{]}perylene molecule ($E$ and $F$ sites
in the main text). Right: $E$ sites having different hypercoordination
number (as indicated), along with their hypercoordinated partners
(white circles). }

\end{figure}
\begin{figure*}
\noindent \centering{}\includegraphics[width=0.9\textwidth]{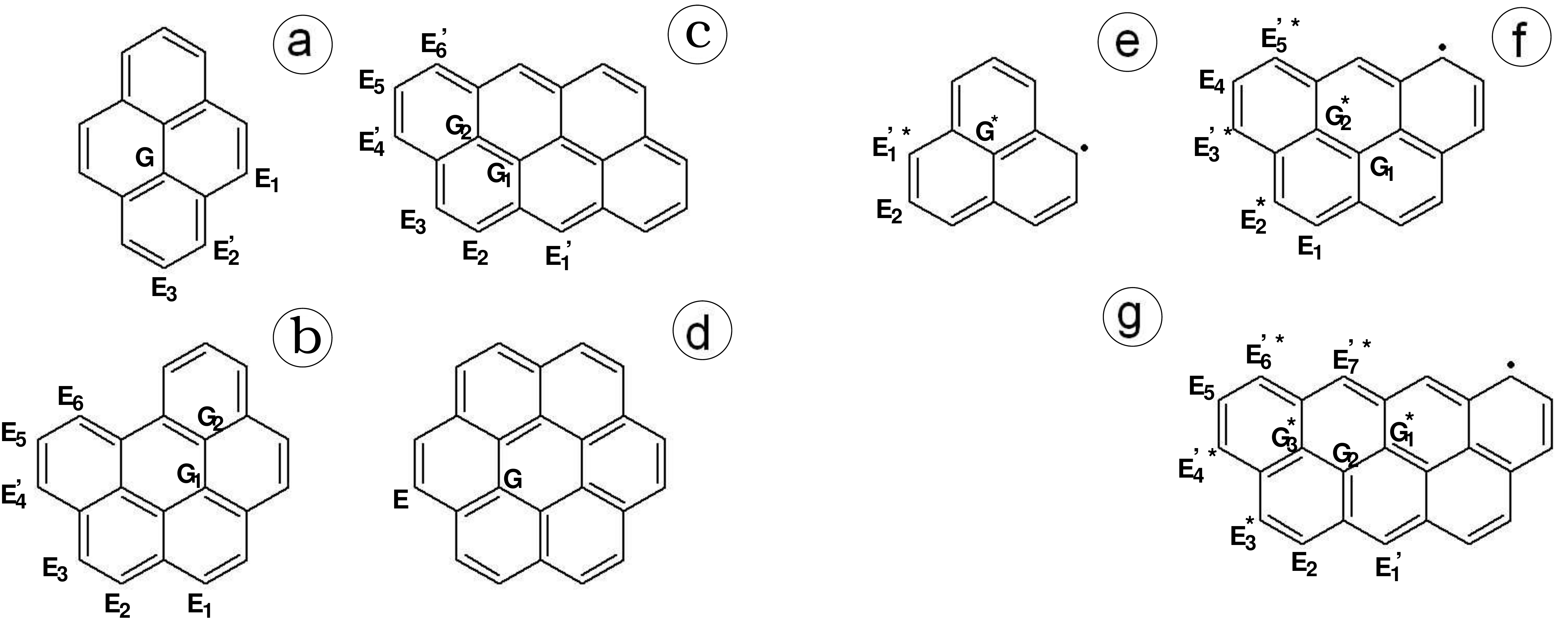}
\caption{\label{fig:PAHs} Balanced (a-d) and imbalanced (e-g) PAHs investigated
in this work, shown \emph{via} one of their possible Lewis structures
(carbon atoms are at the vertexes of the hexagons, and are meant to
be saturated with hydrogen atoms -not shown- if undercoordinated).
(a) Pyrene, (b) anthanthrene, (c) benzo{[}ghi{]}perylene, (d) coronene,
(e) phenalene (peri-naphtene), (f) benzo{[}c{]}pyrene and (g) benzo{[}c{]}anthanthrene.
Also indicated a labeling systems for the adsorption sites considered
in this work, \emph{E} and \emph{G} for {}``edge'' and {}``graphitic''
sites, respectively. A prime is used for edge sites with hypercoodination
number $\xi=2$ and a star is used in (e-g) for the majority sites,
either \emph{E} or \emph{G}, where the unpaired electron (dot) localizes.
See Section \ref{sec:Basic} for details.}

\end{figure*}

\section{\label{sec:Basic}Basic properties of $\pi$ electrons in $sp^{2}$
Carbon structures}

Carbon $sp^{2}$ structures like graphene, GNRs and GDs, are characterized
by a \emph{bipartite} lattice where two distinct sublattices, $A$
and $B$, can be identified such that each $A$ site is connected
to $B$ sites only and \emph{viceversa}. This has important consequences
in the one-electron spectrum if, as it is the case for such structures,
the transfer (hopping) energies beyond the nearest-neighbors are of
secondary importance and the orbital overlap can be neglected. Under
such circumstances, indeed, it is not difficult to prove that the
tight-binding Hamiltonian for the $p_{z}$ orbitals of the $\pi$
electron system has a simple symmetry. Such Hamiltonian reads as \[
H^{TB}=\sum_{<i,j>}t_{ij}a_{i}^{\dagger}b_{j}+h.c.=H_{AB}+H_{AB}\]
where $a_{i}$($a_{i}^{\dagger}$) annihilates (creates) an electron
in site $i$ of the sublattice $A$ (similarly for $b_{j}$($b_{j}^{\dagger}$)
and \textbf{$B$} sublattice sites), $t_{ij}$ is the hopping between
sites $i$ and $j$, and the on-site energy (the energy of an isolated
$p_{z}$ orbital) has been set to zero. Bipartism is responsible for
its (off-)block structure - as emphasized here with the introduction
of $H_{AB}$ and $H_{BA}$ which collectively describe the transitions
$A\rightarrow B$ and $B\rightarrow A$, respectively - and easily
leads to a symmetric spectrum around $\epsilon=0$. The latter is
also the position of the Fermi level with one electron per site (half-filling),
and for this reason the above symmetry is also called electron-hole
symmetry. In conjunction with the spatial symmetry, the presence of
such symmetry is at the origin of the conically shaped band structure
of graphene close to the Fermi level\citep{Intech11}, with interesting
consequences on band-engineering\citep{martinazzo10,casolo10II} and
on the chemical reactivity\citep{casolo09}. Here, to clarify the
connection with chemical reactivity, we focus on some simple results
on the \emph{shape} of low energy (\emph{i.e.} close to the Fermi
level) orbitals that directly follow from such electron-hole symmetry. 

\textbf{Edge localization and hypercoordination}. Low energy orbitals
show a marked tendency to localize on edge sites, as can be easily
seen at the tight-binding level. To this end, we perform a lattice
{}``renormalization''\citep{naumis07,martinazzo10} and focus on
one sublattice only (say $A$) and on the {}``renormalized'' Hamiltonian
$\tilde{H}=H_{AB}H_{BA}$. The renormalized energies $\tilde{\epsilon}_{i}$
are simply related to the eigenvalues%
\footnote{The same holds for eigenvectors, see Ref.\citep{martinazzo10}.%
} $\epsilon_{i}^{\pm}$ of $H^{TB}$, $\epsilon_{i}^{\pm}=\pm\sqrt{\tilde{\epsilon}_{i}}$,
and the renormalized lattice is a triangular lattice (the sublattice
$A$ of the original system) with hopping $t^{2}$ {[}assuming $t_{ij}=t$
for simplicity{]} and on-site energies $t^{2}Z_{i}$, where $Z_{i}$
is the coordination number of the $i-th$ $A$ site in the original
lattice. An edge \emph{necessarily} has undercoordinated ($Z=2$)
sites, hence the ground-state of the renormalized lattice (\emph{i.e.}
the highest occupied/lowest unoccupied molecular orbital {[}HOMO/LUMO{]}
pair of the original lattice) naturally tends to localize on these
sites which present the lowest on-site energy. In the following, we
name $E$ these two-coordinated edge sites, to distinguish them from
those three-coordinated sites which are also present at an edge ($F$
sites), see Fig. \ref{fig:coordination}. Importantly, we expect that
low-energy orbitals localize on $E$ sites and, among these, on those
sites which show the largest number of undercoordinated neighbors
in the renormalized lattice (or, equivalently, next-to-nearest $E$
neighbors in the original lattice) to hybridize with. As is shown
in the following, this latter number turns out to be an important
parameter ruling the reactivity of the edge sites; for this reason
we call it the \emph{hypercoodination} number ($\xi$). Fig. \ref{fig:coordination},
right panel, reports some illustrative cases.

\textbf{Midgap states and spin alignment}. Obviously, energy levels
at $\epsilon=0$, if present, play a major role at half-filling in
determining the reactivity and the magnetic properties of the GDs.
It is instructive to see when this situation occurs, as this also
adds further constraints on the spatial behaviour of the low energy
orbitals. In general, the number of these {}``midgap'' states is
determined by the site-connectivity but their occupancy (spin-alignment)
is solely determined by the \emph{sublattice} \emph{imbalance}. This
follows from a rigorous result proved by Lieb\citep{TheoLieb} for
the realistic (repulsive) Hubbard model having $H^{TB}$ above as
one-electron Hamiltonian: Lieb's theorem states that at half-filling
the ground-state spin $S$ is given by $S=|n_{A}-n_{B}|/2$ where
$n_{A}$ and $n_{B}$ are the number of sites in sublattice $A$ and
$B$, respectively. 

Typically, the number of midgap states matches the sublattice imbalance,
since this is enough to allow for $|n_{A}-n_{B}|$ linearly independent
eigenvectors of $H^{TB}$ at zero energy, all with \emph{null} amplitudes
on the minority sublattice sites\citep{TheoInui}. {[}Accordingly,
in this case, Lieb's theorem above becomes a sort of Hund's rule applied
to the midgap states.{]}. This a simple algebraic result: for, let
$n_{A}>n_{B}$ and $\ket{\psi}=\sum_{i}\alpha_{i}\ket{a_{i}}$ be
a trial solution (here $\ket{a_{i}}=a_{i}^{\dagger}\ket{0}$). At
zero energy, $\sum_{i}\braket{b_{j}|H|a_{i}}\alpha_{i}=0$ must hold
for $j=1,..n_{B}$, which is a set of $n_{B}$ equations for the $n_{A}>n_{B}$
unknowns $\alpha_{i}$ having (at least) $n_{A}-n_{B}$ linearly independent
solutions. This also shows that $\psi$'s \emph{localize on the $A$
lattice sites. }

More generally, the concept of \emph{non}-\emph{adjacent} sites in
a $N$-site bipartite system helps counting the number of midgap states\citep{yazyev10}.
We say that two sites are non-adjacent if they are not bound (connected
by a transfer integral) to each other; for instance, two sites on
the same sublattice are non-adjacent. Clearly, there exists a maximal
set of non-adjacent sites and we call $\alpha$ the sites in this
set, and $\beta$ the remaining ones ($n_{\alpha},n_{\beta}=N-n_{\alpha}$
in number, respectively). Each site $\alpha$ binds at least one site
$\beta$, otherwise it would represent a completely isolated site.
Arranging one electron per site $\alpha$, however, we can form at
most $n_{\beta}$ bonds at a time, and therefore we are left with
$\eta=n_{\alpha}-n_{\beta}=2n_{\alpha}-N$ unpaired electrons. Equivalently,
we end up with $\eta$ midgap states \emph{localized on the maximal
set of non-adjacent sites}. The case of a sublattice imbalance discussed
above is a special result of this rule which, as is evident from the
discussion above, can be equivalently re-phrased by defining $\eta$
to be the number of unpaired electrons in the Lewis structure(s) with
the \emph{maximum} number of $\pi$ (\emph{i.e.} double) bonds.  

We thus see that, in addition to the edge localization discussed above,
depending on the number of sites and their connectivity, there may
exist \emph{topological} constraints which force the carbon $sp^{2}$-system
to have zero energy states. The latter localize on specific lattice
positions which are easily identifiable by inspection. 

\textbf{The systems.} In the following we mainly focus on structures
where the sublattice imbalance is the only source of migdap states,
and call them \emph{balanced} ($S=0$) or \emph{imbalanced} ($S>0$),
accordingly; in particular, only structures with one unit of imbalance
are considered, \emph{i.e.} they all have $S=1/2$, as suggested by
Lieb's theorem and confirmed by our calculations. The considered structures
are shown in Fig.\ref{fig:PAHs}, together with a labeling system
for the sites investigated, which distinguishes the (two-coordinated)
edge sites from the graphitic sites, \emph{E} and \emph{G} in Fig.\ref{fig:PAHs}.
Sites at the edges which are three-coordinated ($F$) are in between
the two categories and will not be considered in the following. With
this exception, all the symmetry unique sites were investigated for
binding of a H atom, with the methods described in the following Section. 

Notice that Fig.\ref{fig:PAHs} further distinguishes those edge sites
which have the largest possible hypercoordination number ($\xi=2$)
with a prime and, where appropriate, identifies with a star the majority
sites (either edge or graphitic) where the midgap states are expected
to localize. As is shown in the following these labels help identifying
the sites with the highest hydrogen affinity (\emph{i.e.} the sites
with the largest binding energy and the smallest barrier to binding).

\section{\label{sec:Computational-methods}Computational methods}

\begin{figure}
\noindent \centering{}\includegraphics[width=0.9\columnwidth]{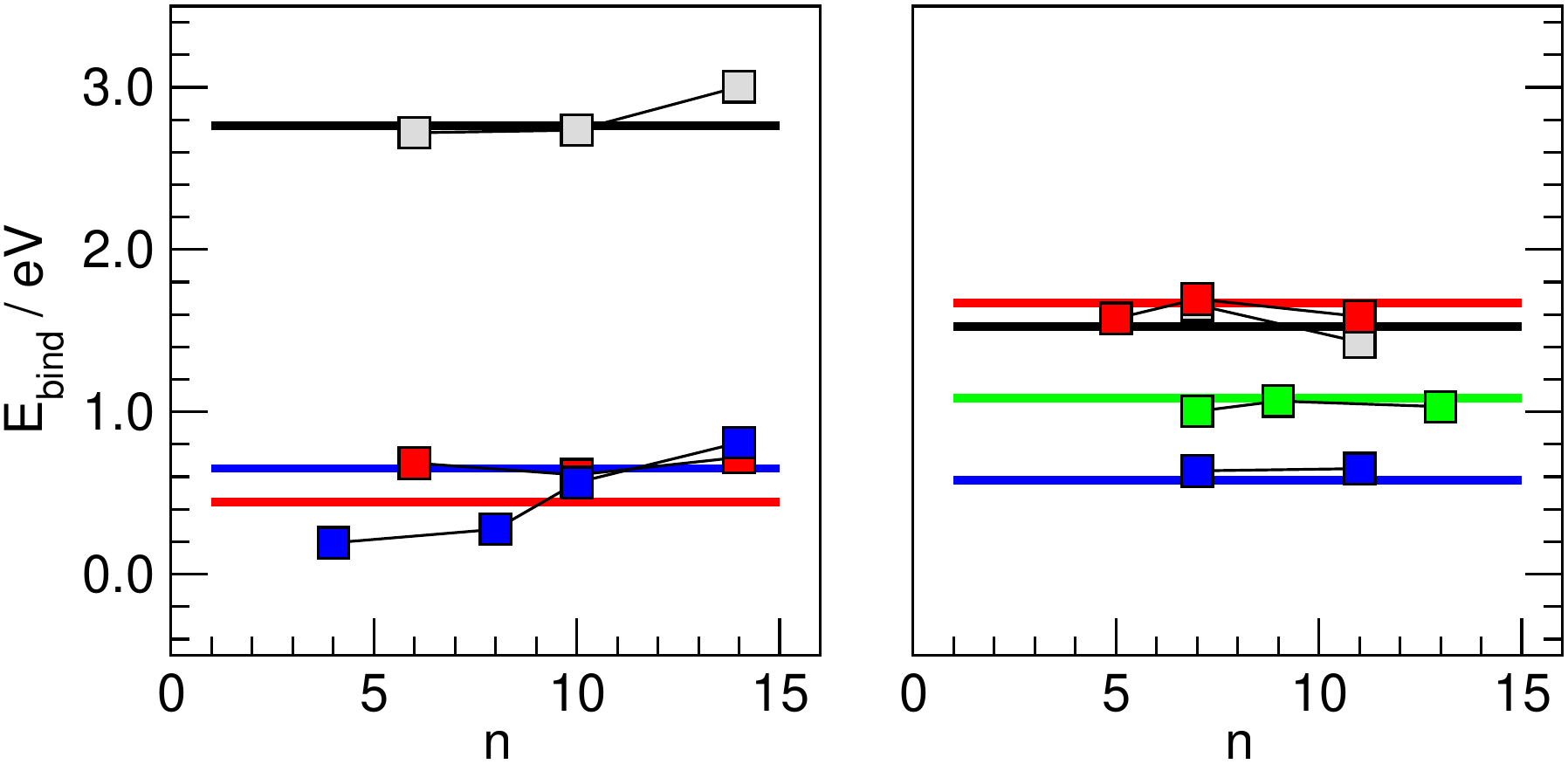}
\caption{\label{fig:convergence-test}Convergence tests on the active space
used in the MCQDPT calculations. The energies for H atom adsorption
are reported for different sites as functions of the number $n$ of
active electrons in a $(n,n)$ correlation scheme. Left: grey, red
and blue symbols for sites $E_{1}$, $E_{2}$ and $G$ of the phenalene
molecule (structure (e) in Fig.\ref{fig:PAHs}). Right: grey, red,
green and blue symbols for sites $E_{1}$, $E_{2}$, $E_{3}$ and
$G$ of pyrene (structure (a) in Fig.\ref{fig:PAHs}). Horizontal
lines mark the values obtained at the DFT level. }

\end{figure}
For each of the selected PAH molecules we computed the binding energy
of a hydrogen atom to the sites labeled in Fig.\ref{fig:PAHs} according
to \[
E_{bind}=E_{\mbox{PAH}}+E_{\mbox{H}}-E_{\mbox{PAH-H}}\]
with two different electronic structure methods. PAH structures were
optimized at the (unrestricted) Density Functional Theory (DFT) level
using the popular B3LYP hybrid exchange-correlation functional with
Dunning's double-valence, atom-centered basis set of the correlation-consistent
type (cc-pVDZ), as implemented in GAUSSIAN 03\citep{g03}. On the
DFT-optimized structures single-point wavefunction calculations were
performed with the same basis-set. These are of the multi-state, multi-reference
perturbation theory type according to the scheme of Hirao\citep{hirao92a,hirao92b,hirao92c,hirao93}
and Nakano\citep{nakano97,nakano98} called Multi-Configuration Quasi-Degenerate
Perturbation Theory (MCQDPT) and implemented in GAMESS\citep{GAMESS08}.
In this scheme dynamical correlation is introduced in a Multi-Configurational
(MC) wavefunction by properly defining a reference one-electron Hamiltonian
based on this wavefunction and computing the second-order perturbation
correction. The chosen MC reference wavefunction was of the Complete
Active Space Self-Consistent-Field (CASSCF) type, where $n$ valence
electrons are distributed in $m$ orbitals (CAS(\emph{n},\emph{m})
in the following) and self-consistency is reached in a variational
optimization. In principle, for the PAHs above a consistent procedure
would require to put all the $\pi$ electrons of the substrate molecules
and that of the H atoms in the same number of orbitals. This is of
course impracticable for all but the smallest molecules, and we therefore
resorted to an orbital localization procedure which takes advantage
of the local character of the bond formation process. We started from
Pipek-Mezey ROHF localized orbitals\citep{PipekMezey} and included
in the active space the $\sigma$ orbital describing the formation
of the $C-H$ bond and the $\pi$ orbitals localized on the sites
which are nearest neighbors of the binding site. This gives rise to
typical CAS(9,9) or CAS(8,8) MCSCF wavefunctions and active spaces
for the perturbation correction. For the smaller PAHs, we performed
some convergence tests on the size of the active space, see Fig.\ref{fig:convergence-test}
for an example. Finally, we also performed plane-wave based, periodic
DFT calculations with the help of the VASP code\citep{VASP1,VASP2},
with parameters similar to those used in our previous works\citep{casolo09,casolo10}
but adapted to a cluster calculation. Briefly, we adopted a $20$
Åx$20$ Åx$20$ Å cell and a 700 eV energy cutoff, with a 1x1x1 $\Gamma$
centered $k$-point grid. Inner electrons were frozen by the projector
augmented wave\citep{PAW1,PAW2} (PAW) approach, and exchange-correlation
effects were handled with the Perdew-Burke-Eznerhof\citep{PBE1} (PBE)
functional in its spin polarized version.

\section{Results and Discussion}

\subsection{Graphitic \emph{vs.} Edge sites}

\begin{figure*}
\noindent \begin{centering}
\includegraphics[width=0.7\textwidth]{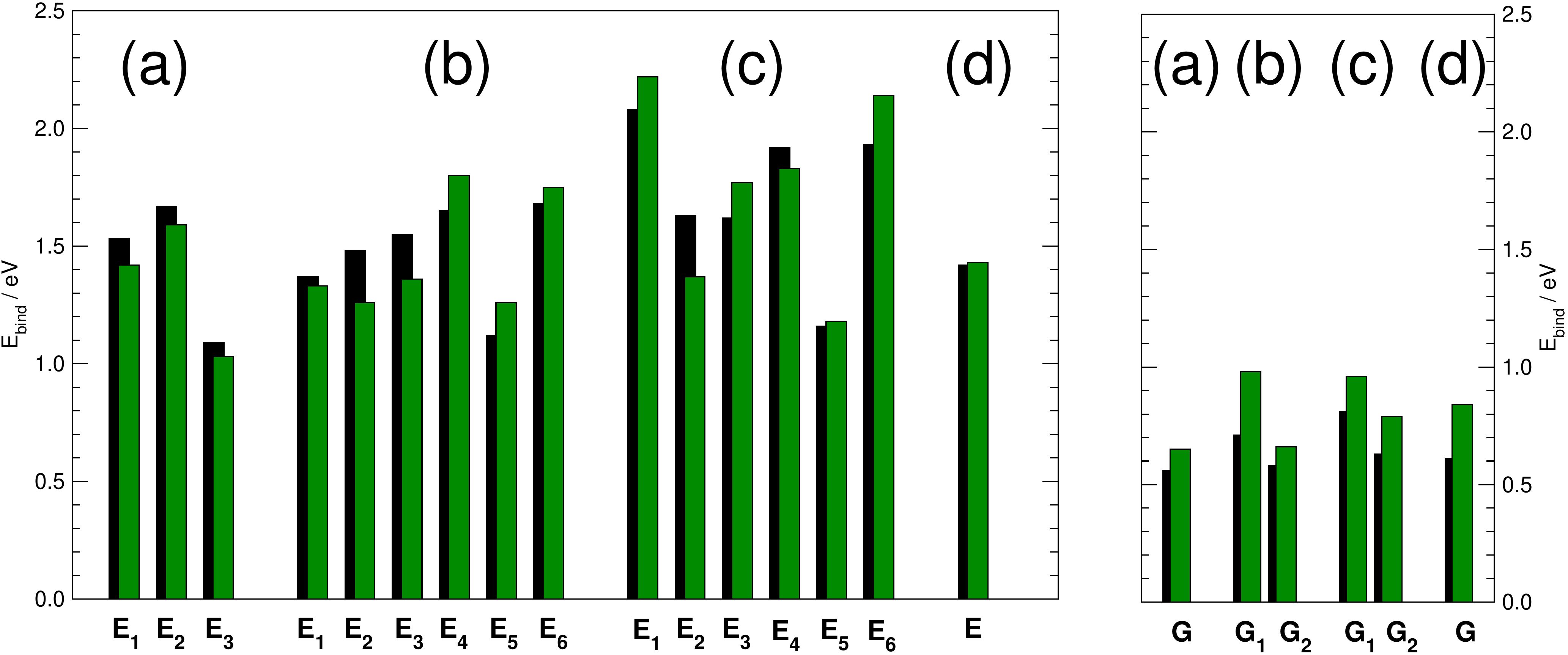}
\par\end{centering}

\caption{\label{fig:balanced_results}Binding energies for \emph{E} (left)
and \emph{G} (right) sites in the structures (a-d) of Fig. \ref{fig:PAHs}.
DFT and MCQDPT results are represented as black and green histograms,
respectively, according to the labeling system of Fig.\ref{fig:PAHs}. }

\end{figure*}
We start by showing the preference for adsorption on the edge sites
which was already noted by several authors\citep{Bauschlicher98,zwier10,HornekaerAstro08,Ahlrichs00,hammer11}.
Fig. \ref{fig:balanced_results} shows the computed binding energy
for all the \emph{E} and \emph{G} sites of the structures (a-d) of
Fig. \ref{fig:PAHs}, as obtained in the ground-state spin manifold
of the total system%
\footnote{For DFT calculations we refer here to the spin of the Kohn-Sham non-interacting
determinant. Spin contamination is always found minimal, \emph{i.e.}
the determinants are close to be eigenstates of the total (squared)
spin operator, apart from having a well defined projection.%
}, $S=1/2$. The DFT results (black histograms) compare very well with
the available literature data. For instance, for the pyrene molecule
we find 1.53, 1.67 and 1.09 eV for the sites $E_{1}$, $E_{2}$ and
$E_{3}$ which compare well with the values 1.50, 1.61 and 1.06 eV
recently obtained by Rasmussen \emph{et al.}\citep{hammer11} with
a real-space implementation of the DFT-PBE level of theory. 

Clearly, a striking difference between \emph{E} and \emph{G} sites
is apparent from Fig. \ref{fig:balanced_results}: binding energies
at an \emph{E} site can be as large as twice the binding energy for
a \emph{G} site. The latter, on the other hand, compare rather well
with the value of the hydrogen atom adsorption energy in graphene\citep{Allouche02,Allouche2005,casolo09}
and graphite\citep{Sha&Jackson2002,Allouche2005}. This simple finding,
together with a corresponding behaviour for barrier energies to be
discussed below, already suggests that the edges of realistic samples
could be active sites where hydrogenation starts and propagates into
the bulk.

\subsection{Geometric \emph{vs.} electronic effects}

\begin{figure}
\noindent \begin{centering}
\includegraphics[clip,width=0.95\columnwidth]{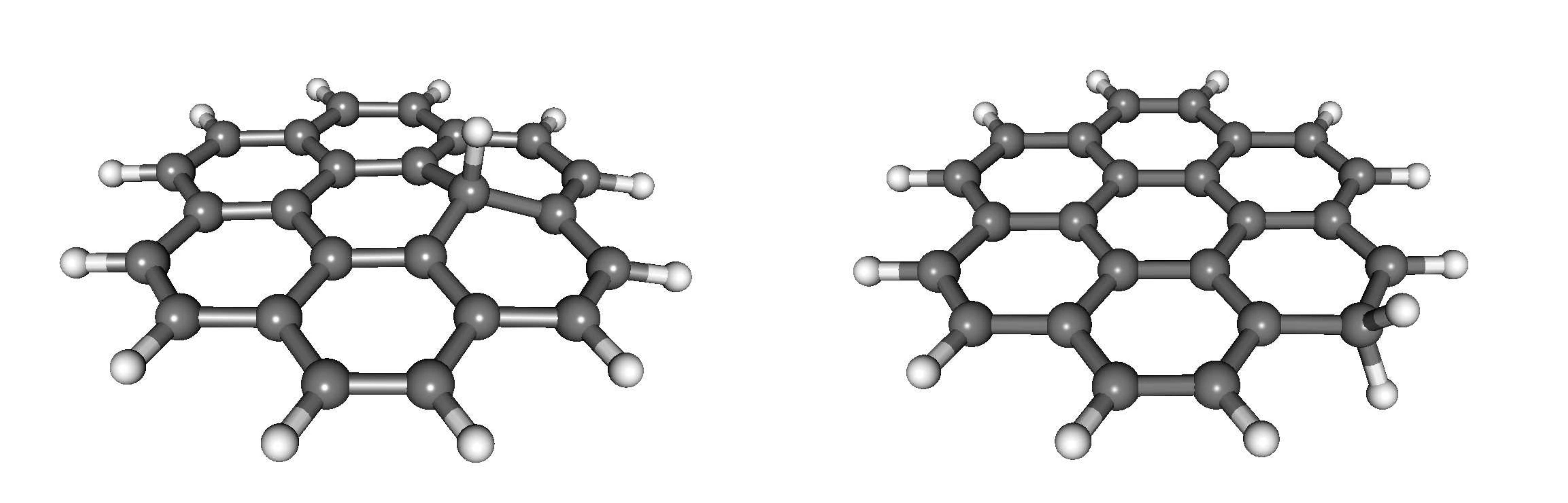}(a)
\par\end{centering}

\noindent \centering{}\includegraphics[clip,width=0.6\columnwidth]{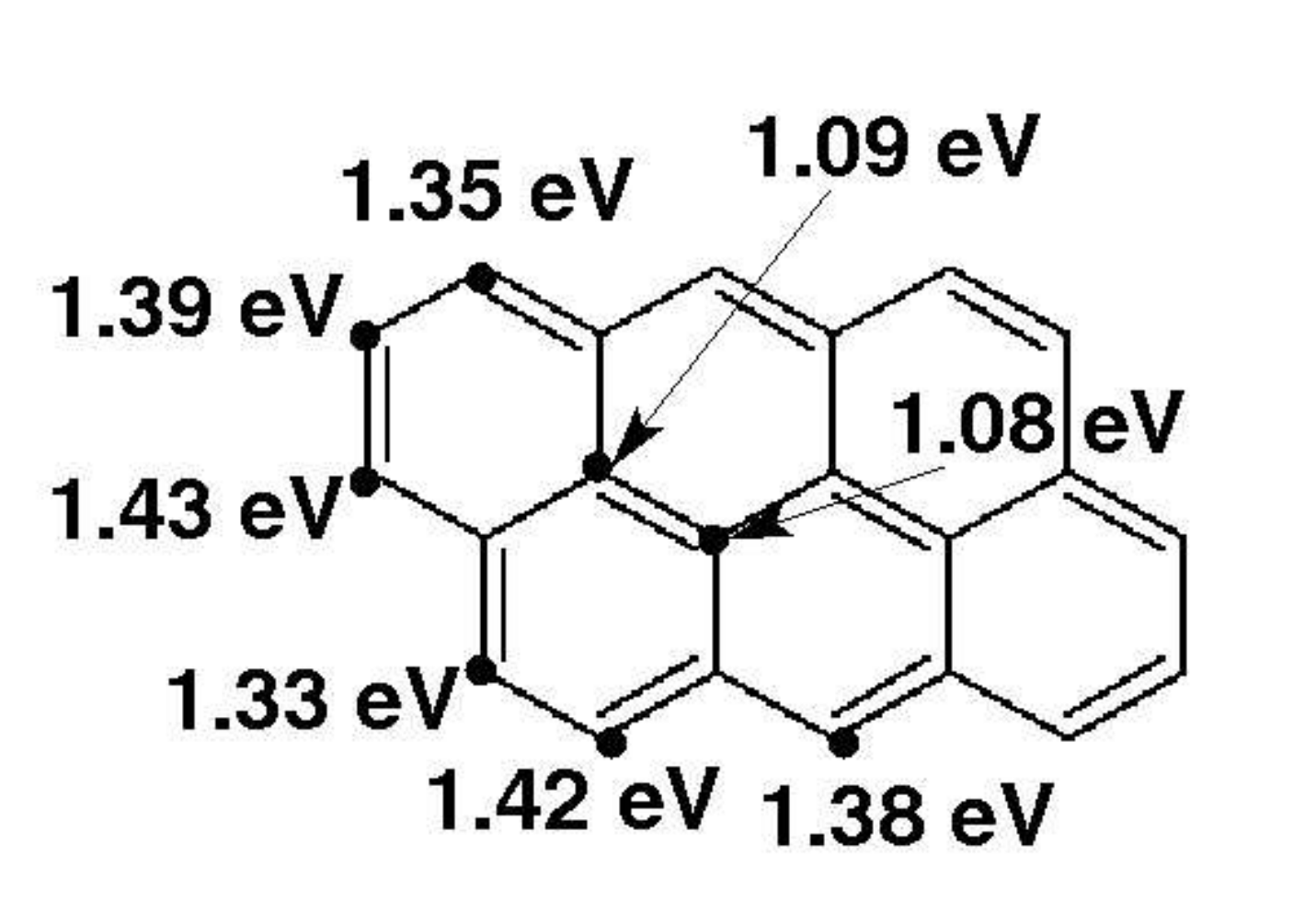}(b)
\caption{\label{fig:structures}(a) Optimized structures for the singly hydrogenated
coronene molecule. Left and right panel for adsorption on the \emph{G}
and \emph{E} site, respectively. (b) Reorganization energy for adsorption
of a H atom in the indicated sites of the benzo{[}ghi{]}perlyene molecule.}

\end{figure}
Before analysing the results in details, we show here that {}``geometrical''
effects \emph{per se} cannot explain the different behaviour of edge
and inner sites evident in Fig.\ref{fig:balanced_results}. Binding
of a H atom on a $sp^{2}-$carbon atom requires a $sp^{2}\rightarrow sp^{3}$
rehybridization which leads to a tetrahedral reorganization of the
bonding partners, as is shown in Fig. \ref{fig:structures} (a) for
the case of the coronene molecule. Without such re-arrangement of
the local environment no binding would occur: a local substrate relaxation
is essential to {}``prepare'' the electronic structure for binding,
but this too is affected by the overall electronic structure which
is always dominated by molecular orbitals spreading all over the molecule. 

A simple (but wrong) argument would suggest that the \emph{same} local
re-arrangement which occurs upon bonding (but \emph{without} the H
{}``probe'') requires \emph{less} energy for an edge than for an
inner carbon atom, since in the first case at least one of the bonding
partners is a monovalent species not embedded in the molecular network.
We can define this \emph{reorganization} energy as \[
E_{R}=E_{eq}^{*}(PAH)-E_{eq}(PAH)\]
where $E_{eq}(PAH)$ is the energy of the pristine molecule in the
equilibrium configuration and $E_{eq}^{*}(PAH)$ is the energy of
the molecule in the same distorted configuration that it takes when
binding the H atom. In contrast to the expectation above, we find
that $E_{R}$ for an \emph{E} site is always \emph{larger} than that
for a \emph{G} site. For coronene, for instance, we obtain 1.40 eV
and 1.04 eV, respectively, at the DFT level of theory, and similar
values are found for all the structures considered in this work: the
reorganization energy is $\sim$1.4$\pm$0.1 eV for \emph{E} sites
and $\sim$1.0$\pm$0.1 eV for \emph{G} sites, see for instance Fig.
\ref{fig:structures}(b) for the case of the benzo{[}ghi{]}perylene.
We thus see that the preference in binding a H atom to an edge site
occurs \emph{despite} the larger reorganization energy needed at these
kind of sites. This allows us to conclude that this preference is
due to the electronic effects introduced in Section \ref{sec:Basic}.

\subsection{Hypercoordination}

\begin{figure}
\noindent \centering{\includegraphics[clip,width=0.7\columnwidth]{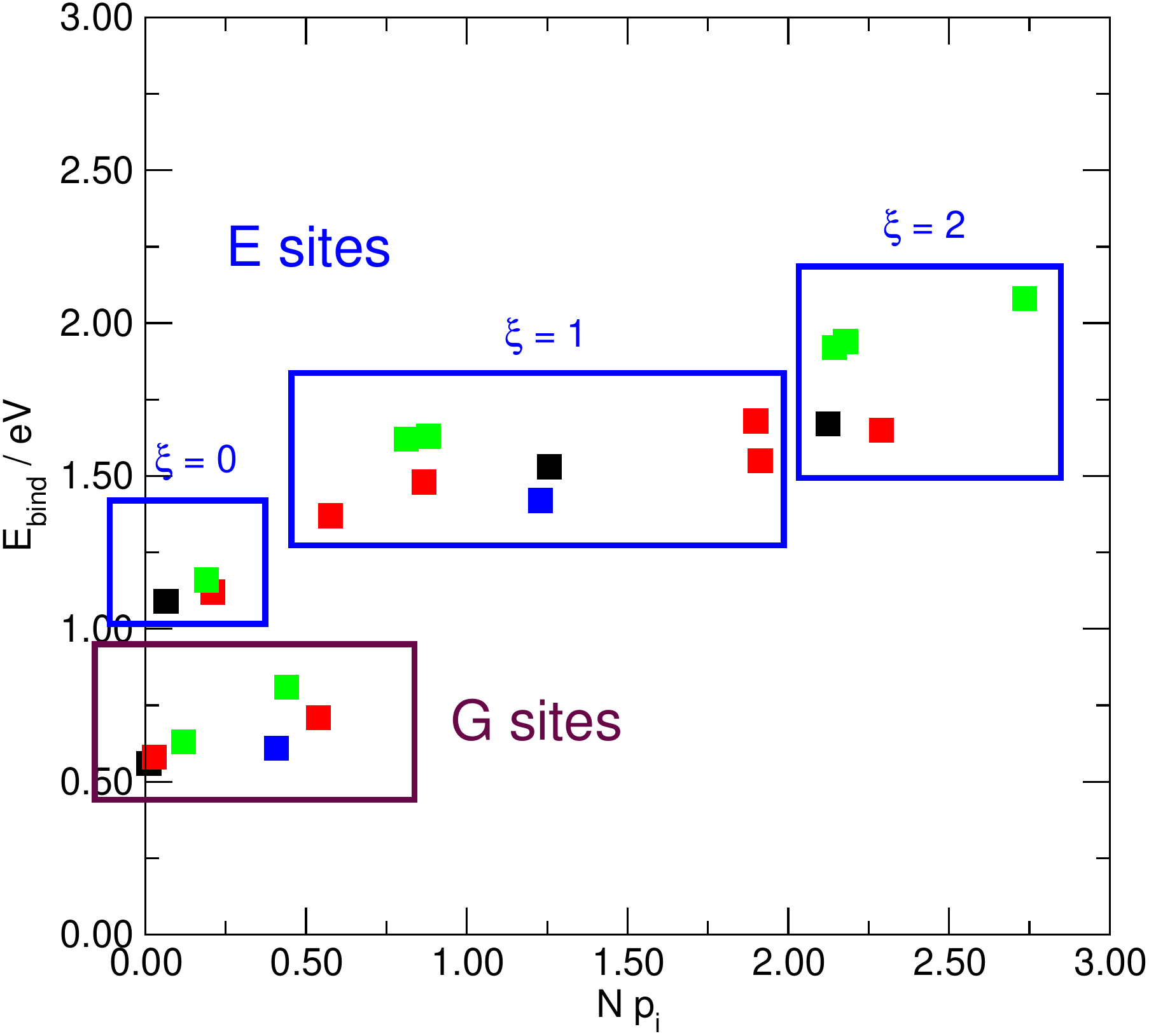}}
\caption{\label{fig:Hypercoord}Binding energies as functions of the site normalized
populations of the substrate HOMO. Black, red, green and blue symbols
for structures (a-d), respectively. Also indicated the hypercoordination
number of the edge sites. }

\end{figure}
Next we discuss the results of Fig. \ref{fig:balanced_results} in
detail since, apart from the overall behaviour, the binding energies
can take quite different values depending on the site they refer to.
A closer inspection reveals that the values for the interesting \emph{E}
sites correlate very well with the hypercoordination number introduced
in Section \ref{sec:Basic}: the lager is the hypercoordination the
larger is the binding energy. This can be made evident by reporting
the results of Fig. \ref{fig:balanced_results} as functions of the
site populations $p_{i}$ of the HOMO; the latter are meant here per
spin species, and were obtained by a Mulliken analysis of the molecular
orbitals of the pristine molecules, as computed with a restricted
Kohn-Sham determinant. This is shown in Fig.\ref{fig:Hypercoord}
where the populations have been normalized to the values they would
have if the HOMOs spread  over all carbon atoms ($1/N$). Fig. \ref{fig:Hypercoord}
shows that the binding energies correlate well with $Np_{i}$. The
trend is roughly linear and different for the \emph{E} and the \emph{G}
sites, but we did not attempt to extract any behaviour because of
the limited number of data available. More importantly, Fig. \ref{fig:Hypercoord}
shows that the energies correlate well with the hypercoordination
number $\xi$ of the site, particularly if the comparison is made
between sites of the same molecule. As already emphasized above, this
number can be readily obtained by simply inspecting the carbon structure
under study.

\subsection{Imbalanced structures}

\begin{figure*}
\begin{centering}
\includegraphics[clip,width=0.7\textwidth]{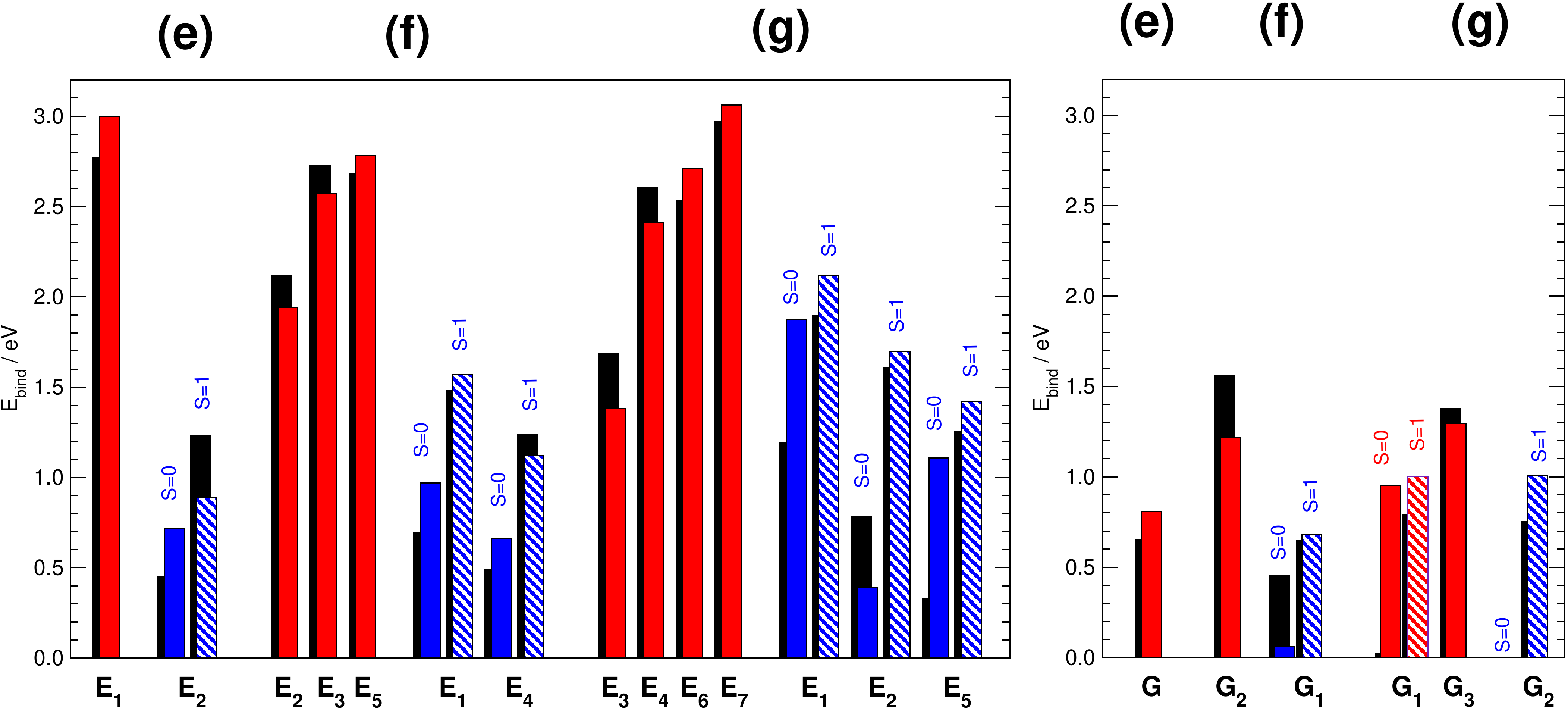}
\par\end{centering}

\caption{\label{fig:imbalanced}Binding energies for \emph{E} (left) and \emph{G}
(right) sites in the structures (e-g) of Fig. \ref{fig:PAHs}. DFT
and MCQDPT results are represented as black and colored histograms,
respectively, according to the labeling system of Fig.\ref{fig:PAHs}.
The color code (for MCQDPT results only) is red for majority and blue
for minority sites; occasionally, both the singlet (full color) and
the triplet (shaded) spin manifolds have been considered, as indicated.
See text for details.}

\end{figure*}
Next we move to the more complicated situation (the doublet structures
(e-g) of Fig.\ref{fig:PAHs})) where topological constraints lead
to the appearance of zero-energy modes and additional {}``localization''.
Analogously to the results of Fig. \ref{fig:balanced_results}, we
find also in this case a clear distinction between edge and graphitic
sites. This is evident from Fig. \ref{fig:imbalanced} where we report
the binding energies for the structures (e-g) on a larger energy scale
than the one used in Fig. \ref{fig:balanced_results}. This is one
of the consequences of the additional electronic effect due to the
appearance of the (singly occupied) midgap state: binding of two radical
species only requires coupling of their unpaired electrons and is
thus typically much more energetic than in the case where a bond has
to be broken. A further consequence is a rough splitting of the results
into two {}``branches'', according to whether the relevant site
belongs or not to the majority set (red and blue blocks of results
in Fig. \ref{fig:imbalanced}). Notice that, according to Lieb's theorem,
in the first case the resulting total spin state is a singlet, whereas
in the second case is a triplet. This indeed what we find: Fig.\ref{fig:imbalanced}
shows that for adsorption of a H atom on a minority site the binding
energy in the triplet state is larger than in the singlet. Not shown
in the figure, we also checked that adsorption on a majority site
occurs more favourably in the singlet manifold; this is true for all
cases considered but the site $G_{1}$ of structure (g) where we find
that H binds more favourably in the triplet state%
\footnote{This is likely due to the importance of next-to-nearest neighbor hing
hoppings which are implicitly included in the \emph{ab-initio} calculations
and are enough to invalidate Lieb's theorem. Notice that in this way
the resulting chemical structure keeps unaltered two naphtalene moieties. %
}. 

\begin{figure}
\noindent \centering{\includegraphics[clip,width=0.9\columnwidth]{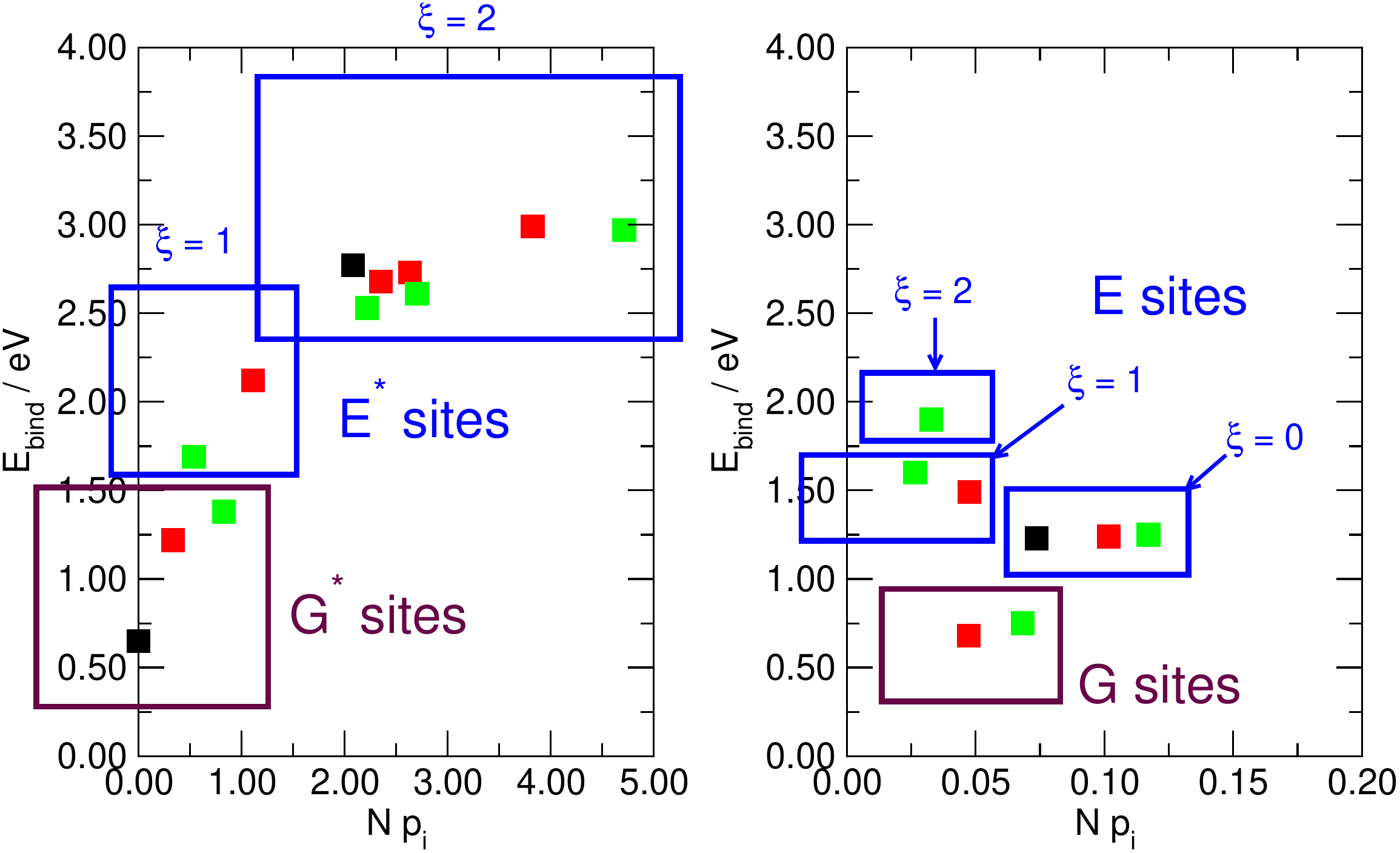}}
\caption{\label{fig:Hypercoord-2}Binding energies as functions of the site
normalized populations of the substrate HOMO. Black, red and green
symbols for structures (e-g), respectively. Also indicated the hypercoordination
number of the edge sites. Left and right panels for majority and minority
sites, respectively. }

\end{figure}

We thus see that, in the case considered in this section, the energy
ordering arises from the complicated interplay between coordination,
hypercoordination and topological frustration. For this reason, in
plotting the results as functions of the normalized populations, analogously
to Fig. \ref{fig:Hypercoord}, we consider separately the majority
and the minority sites, reported in the left and right panels of Fig.
\ref{fig:Hypercoord-2}, respectively. We see now that a good correlation
between the binding energies and the HOMO populations is found only
for the majority sites, nevertheless the hypercoordination number
remains a good parameter for establishing the right energy ordering
within each category: the binding energy is found to monotonically
increase when increasing $\xi$. In general, majority sites show larger
binding energies of minority sites with the same coordination number
(\emph{i.e.} either \emph{E} or \emph{G}) but, even for the same molecule,
a large hypercoordination may offset the topological frustration of
a minority site. For instance, the (minority) site $E_{1}$ in structure
(g) shows a larger binding energy than the (majority) site $E_{3}$;
notice though that the {}``expected'' ordering is restored if comparison
is made between results for the same spin manifold. In general, however,
the most favoured (relevant) final hydrogenated structures are always
easily identified: they are obtained by binding a H atom to the majority
\emph{E} sites with the largest hypercoordination number. %

Notice further that imbalanced structures also arise after a H atom
has been adsorbed onto any of the balanced structures (a-d), since
formation of a CH bond effectively removes one carbon $p_{z}$ orbital
from the $\pi$ network and thus acts as a vacancy. In this case,
hydrogen bonding to form a \emph{dimer} follows the same rules. For
instance, Rauls and Hornekaer \citep{HornekaerAstro08} used DFT-PW91
to systematically investigate hydrogenation of coronene up to saturation.
They found that addition of a H atom to the most stable H-coronene
structure (\emph{i.e.} with a first H bound to a \emph{E} site) is
most favoured in the \emph{ortho}-edge position, \emph{i.e.} on the
\emph{E} site which is nearest neighbor to the first adsorption site.
This is a majority site with an effective coordination number $Z=1$,
which would correspond to an additional type of site, {}``\emph{D}''.
Furthermore, five \emph{E} sites exist in H-coronene with $\xi=1$
having a large binding energy. Analogous results holds for pyrene,
see Rasmussen \emph{et al.}\citep{hammer11}.

\subsection{Adsorption profiles}

\begin{figure}
\noindent \begin{centering}
\includegraphics[clip,width=1\columnwidth]{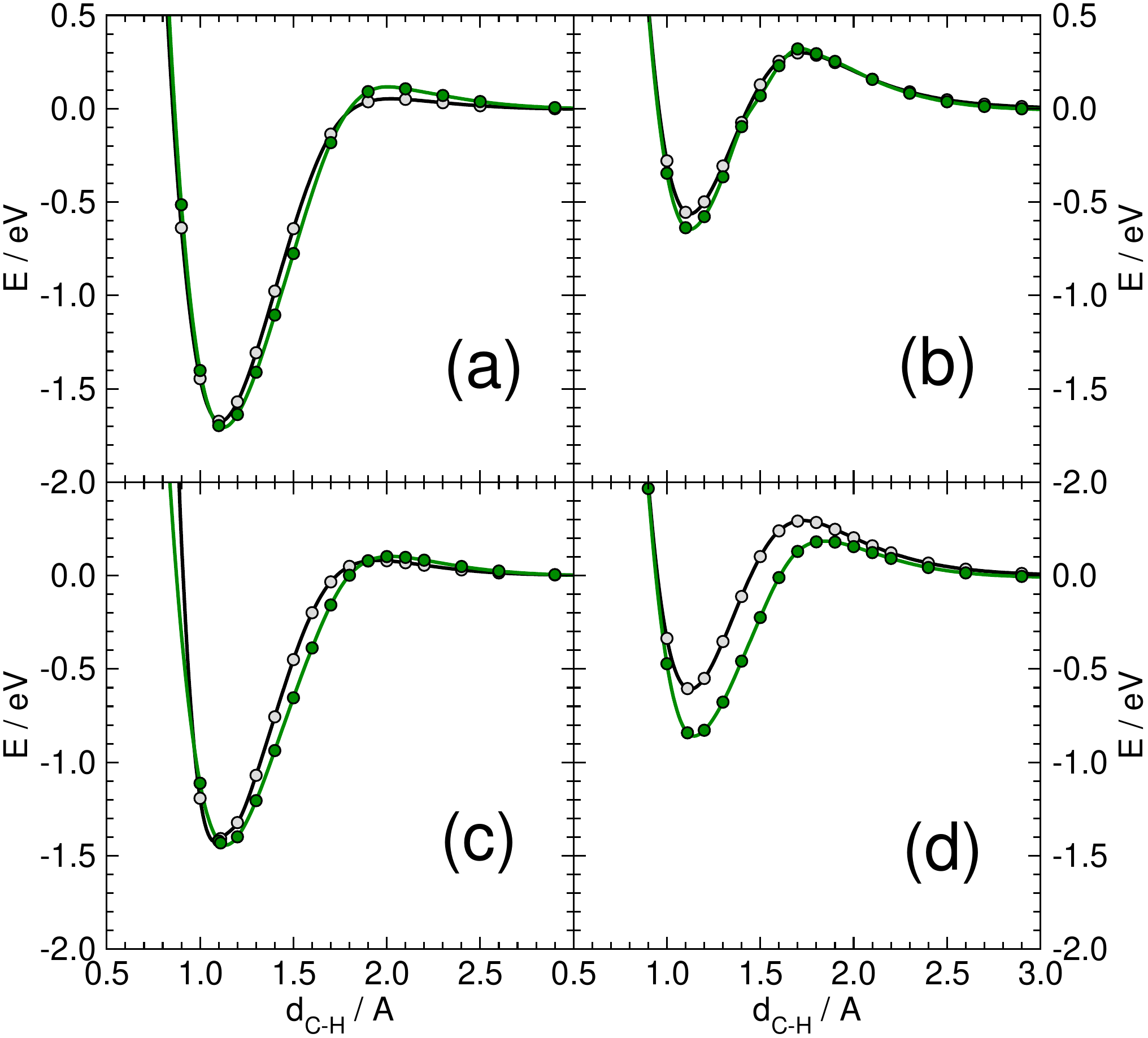}
\par\end{centering}

\caption{\label{fig:paths}Hydrogen adsorption paths on pyrene (a-b) and coronene
(c-d), on the left for an edge site and on the right for a graphitic
site. Black and green symbols for DFT and MCQDPT results. Lines are
spline interpolation to guide the eyes. }

\end{figure}
We now look at the full energy profiles (minimum energy paths) for
a H atom adsorption, focusing on few illustrative cases. We show in
particular that the arguments used so far for the adsorption energies
equally apply to the energy \emph{barriers} for the H atom sticking.
Thus, the energy ordering rules drawn in the previous sections not
only determines the \emph{thermodynamics} but also the \emph{kinetics}
of the hydrogenation process. 

Hydrogen atom binding is an activated process with an energy barrier
which typically prevents adsorption under room temperature conditions\citep{Hornekaer2006,Hornekaer2006a}.
For instance, in graphite (graphene) the barrier is $\sim$0.2 eV
high and this prevented for some time observation of a chemisorbed
hydrogen phase. This barrier is typically linearly related to the
binding energy itself\citep{casolo09}, in accordance with the general
finding (known as Brønsted\textendash{}Evans\textendash{}Polayni rule)
that a larger reaction exothermicity is accompanied by a lower energy
barrier. The same applies here, as is shown for the cases of pyrene
and coronene reported in Fig.\ref{fig:paths}, for both an \emph{E}
and a \emph{G} site. Such curves have been obtained by fixing the
CH distance at the desired value and performing a full structural
relaxation of the remaining degrees of freedom at the DFT-B3LYP level
of theory. As is evident from the figure, a larger binding energy
reflects a smaller adsorption barrier, which can be even almost vanishing
when H binding occurs at an edge site. Similar results hold for all
the paths considered in this work, \emph{i.e.} for H atom adsorption
on most of the sites considered in Fig.\ref{fig:PAHs}. As already
noticed above this finding suggests that the edges of realistic samples
could be active sites where hydrogenation starts and propagates into
the bulk: addition of H atoms to \emph{E} sites modifies the sublattice
imbalance and at the same time effectively converts a number of \emph{F}
(\emph{G}) sites into \emph{E} (\emph{F}) sites.

\subsection{Correlation level }

\begin{figure}
\noindent \begin{centering}
\includegraphics[width=1\columnwidth]{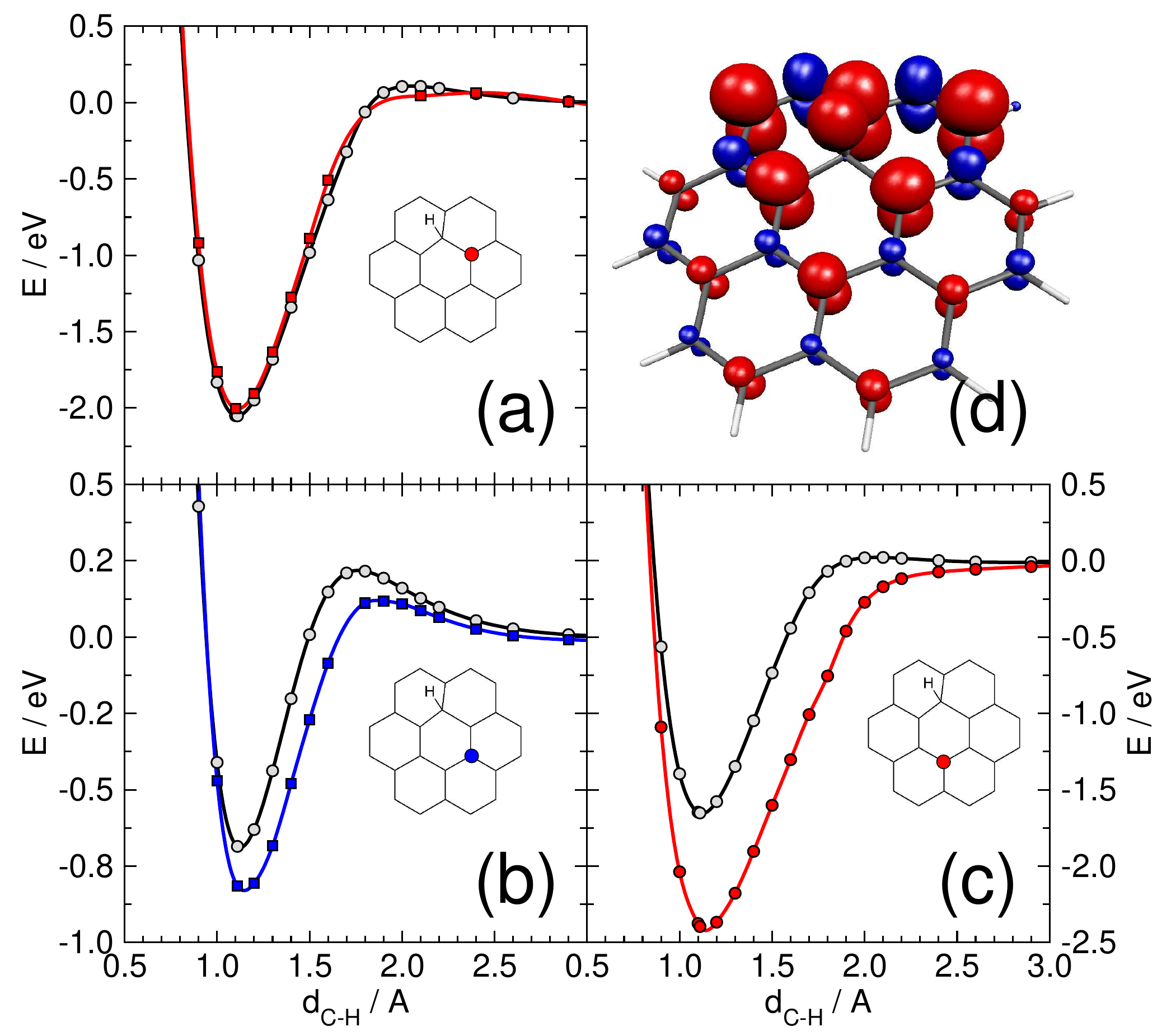}
\par\end{centering}

\caption{\label{fig:coroH2}Adsorption curves for a second H atom in the (graphitic)
\emph{ortho}- (a), \emph{meta}- (b) and \emph{para}- (c) positions
with respect to a first H atom, as indicated in the insets. Grey and
colored symbols for DFT and MCQDPT results, respectively, and lines
are spline interpolation to the data for guiding the eyes. Panel (d)
shows the spin-density of the H-coronene substrate with a H atom adsorbed
on a \emph{G} site. }

\end{figure}
Finally, we focus on some technical aspects concerning the treatment
of electron correlation. Though not emphasized so far, the results
of the DFT-B3LYP calculations have been shown in parallel to the results
of more accurate, though more expensive, MCQDPT calculations (see
Section \ref{sec:Computational-methods}) which we performed on the
DFT-optimized structures. As is evident from Fig.s \ref{fig:balanced_results},\ref{fig:imbalanced}
and \ref{fig:paths} the two sets of data agree well with each other,
the discrepancies being at most few tenths of eV in few cases. No
general trend is found in the comparison, except maybe for a general
tendency of the correlated wavefunction calculations to give a larger
binding energy than DFT for the graphitic sites, see \emph{e.g.} the
right panel of Fig. \ref{fig:balanced_results}. This is particularly
evident for the \emph{G} site of the coronene molecule: Fig. \ref{fig:paths}
(d) shows that binding to this site is $\sim$0.2 eV stronger when
computed at the MCQDPT than at the DFT level of theory, and that a
corresponding trend is found for the barrier. However, given the limited
number of active electrons that could be consistently included in
the wavefunction calculations we doubt that this discrepancy is a
manifestation of a true physical effect. This is made more evident
in Fig. \ref{fig:coroH2}, where the adsorption paths for a second
H atom onto the \emph{ortho}-, \emph{meta}- and \emph{para}- position
to the first \emph{G} sites are displayed for the two different levels
of theory. We chose to focus on this system because of the role it
played as a cluster model for graphene (graphite) since Jeloaca and
Sidis\citep{Jeloaica1999} used it to investigate H atom adsorption
on the graphitic sites. As is clear from Fig. \ref{fig:coroH2} the
above discrepancy doubles when adsorption proceeds in \emph{para}-
but vanishes for the \emph{ortho}- site, thereby suggesting that the
{}``extension'' of the structure may be a source of error in the
MCQDPT calculations. Notice that the wavefunction calculations are
always \emph{two}-\emph{state} MCQDPT calculations, in order to correctly
handle the barrier region, and included in some cases a level shift
correction to get rid of the intruder state problem. 

Finally, we performed few additional calculations of the binding energies
with a very different implementation of the DFT-GGA theory, namely
a $\Gamma$-point, periodic plane-wave calculation using a pure GGA
functional as described in Section \ref{sec:Computational-methods}.
We find for coronene 1.42 and 0.67 eV for adsorption on the \emph{E}
and the \emph{G} site, respectively, which compare very well with
the values obtained with the hybrid B3LYP functional, namely 1.42
and 0.61 eV. The same holds for the adsorption of a second atom on
the same sites considered in Fig. \ref{fig:coroH2}: we obtain 2.04,
0.70 and 1.82 eV for the \emph{ortho}-, \emph{meta}- and \emph{para}-
graphitic sites, to be compared with 2.05, 0.69 and 1.65 eV. Notice
that also in this case the larger discrepancy occurs at the \emph{para}-
position, which might signal the need of additional care in the correlation
problem.

\section{Summary and Conclusions}

We considered atomic hydrogen adsorption on a number of small graphenic
structures (PAH molecules) in order to investigate the enhanced reactivity
of the edge sites already observed by several authors. To this end,
we selected only small structures to prevent the formation of radical
species at the edge, as it occurs with the formation of zero-energy
states at the edges of large zig-zag nanoribbons. Surprisingly, we
found that some edge localization always occurs as a consequence of
the reduced coordination of \emph{E} sites which translates into a
lower on-site energy in a renormalized lattice. Further localization
occurs when \emph{E} sites are highly coordinated in the renormalized
lattice, as measured by a {}``hypercoordination'' number $\xi$.
We found a very good correlation between the binding (barrier) energies
and the coordination and hypercoordination numbers: the most favoured
sites for H atom adsorption (but likely for adsorption of any monovalent
species used to form covalent bonds with carbon) are those showing
the lowest coordination and the largest hypercoordination numbers
($E'$ sites in Fig. \ref{fig:PAHs}). We also found, similarly to
graphene, that further enhancement of the reactivity of specific lattice
positions may arise from the same topological frustration which gives
rise to midgap states, \emph{i.e.} that occurring when the maximal
set of non-adjacent sites exceeds half the total number of sites.
In this case a preference towards the maximal set of non-adjacent
sites adds to the above preference for coordination and high hypercoordination. 

We obtained these results in small (subnanometer-sized) graphene structures,
but they are expected to hold for more complex structures. For instance,
hydrogenation is known to occurs much more easily on a zig-zag than
on an armchair edge of large area graphene: May \emph{et al.} \citep{Ahlrichs00},
for instance, extrapolated DFT values computed on finite size graphenes
towards the infinite size limit and obtained 2.86$\pm$0.15 eV for
the zig-zag edge and 1.74$\pm$0.11 eV for the armchair one. This
is consistent with the {}``rules'' found here. Indeed, both edges
have \emph{F} and \emph{E} sites, but only zig-zag \emph{E} sites
can be fully hypercoordinated: $\xi=2$ in this case, to be compared
with $\xi=0$ for the \emph{E} sites of an armchair edge. 

Beside their simplicity, one of the main advantage of the derived
rules is that they are based on \emph{local} considerations which
hold irrespective of the global electronic properties of the carbon
nanostructure under study. As a consequence, our findings suggest
that exposing arbitrarly shaped graphene dots to controlled amount
of atomic hydrogen (\emph{e.g.} under cold plasma conditions) hydrogenation
starts from the edges and propagates into the bulk in a much more
efficient way than expected solely on the basis of the bulk adsorption
energetics. 

\bibliographystyle{jcp}

\end{document}